\documentclass[9pt,twocolumn,twoside]{revtex4}

\usepackage{amsfonts,amsmath,amssymb,bm,graphicx,color,times,gensymb,setspace} 

\newcommand{\scgo}{SrCoGe$_2$O$_6$}


\begin{document}

\title{Cobalt-Based Pyroxenes: A New Playground for Kitaev Physics}

\author{Pavel Maksimov}
\affiliation{Bogolyubov Laboratory of Theoretical Physics, Joint Institute for Nuclear Research, Dubna, Moscow region 141980, Russia}
\affiliation{M.N. Mikheev Institute of Metal Physics UB RAS, 620137, S. Kovalevskaya str. 18, Ekaterinburg, Russia}

\author{Alexey V. Ushakov}
\affiliation{M.N. Mikheev Institute of Metal Physics UB RAS, 620137, S. Kovalevskaya str. 18, Ekaterinburg, Russia}

\author{Andey F. Gubkin}
\affiliation{M.N. Mikheev Institute of Metal Physics UB RAS, 620137, S. Kovalevskaya str. 18, Ekaterinburg, Russia}
\affiliation{Institute of Natural Sciences and Mathematics, Ural Federal University, Mira St. 19, 620002 Ekaterinburg, Russia}

\author{G\"unther J. Redhammer}
\affiliation{Department of Chemistry and Physics of Materials, University of Salzburg, Jakob-Haringer-Strasse 2a, Salzburg A-5020, Austria}

\author{Stephen M. Winter}
\affiliation{Department of Physics and Center for Functional Materials, Wake Forest University, NC 27109, USA}

\author{Alexander I. Kolesnikov}
\affiliation{Neutron Scattering Division, Oak Ridge National Laboratory, Oak Ridge, TN 37831, USA}

\author{Antonio M. dos Santos}
\affiliation{Neutron Scattering Division, Oak Ridge National Laboratory, Oak Ridge, TN 37831, USA}

\author{Zheng Gai}
\affiliation{Center for Nanophase Materials Sciences, Oak Ridge National Laboratory, Oak Ridge, TN 37831, USA}

\author{Michael A. McGuire}
\affiliation{Materials Science and Technology Division, Oak Ridge National Laboratory, Oak Ridge, TN 37831, USA}

\author{Andrey Podlesnyak}
\affiliation{Neutron Scattering Division, Oak Ridge National Laboratory, Oak Ridge, TN 37831, USA}

\author{Sergey V. Streltsov}
\affiliation{M.N. Mikheev Institute of Metal Physics UB RAS, 620137, S. Kovalevskaya str. 18, Ekaterinburg, Russia}
\affiliation{Department of Theoretical Physics and Applied Mathematics, Ural Federal University, Mira St. 19, 620002 Ekaterinburg, Russia}

\begin{abstract}
Recent advances in the study of cobaltites have unveiled their potential as a promising platform for realizing Kitaev physics in honeycomb systems and the Ising model in weakly coupled chain materials. In this manuscript, we explore the magnetic properties of pyroxene SrCoGe$_2$O$_6$ using a combination of neutron scattering, {\it ab initio} methods, and linear spin-wave theory. Through careful examination of inelastic neutron scattering powder spectra, we propose a modified Kitaev model to accurately describe the twisted chains of edge-sharing octahedra surrounding Co$^{2+}$ ions. The extended Kitaev-Heisenberg model, including a significant anisotropic bond-dependent exchange term with $K/|J|=0.96$, is identified as the key descriptor of the magnetic interactions in SrCoGe$_2$O$_6$. Furthermore, our heat capacity measurements reveal an effect of an external magnetic field (approximately 13~T) which shifts the system from a fragile antiferromagnetic ordering with $T_{\mathrm{N}}=9$~K to a field-induced state. We argue that pyroxenes, particularly those modified by substituting Ge with Si and its less extended $p$ orbitals, emerge as a novel platform for the Kitaev model. This opens up possibilities for advancing our understanding of Kitaev physics.
\end{abstract}

\date{\today}

\maketitle


Kitaev materials, where strong spin-orbit coupling (SOC) results in a pronounced exchange anisotropy frustrating a magnetic sub-system, have become one of the most actively studied subjects in physics of magnetic materials~\cite{Trebst2017,Takagi2019}. This is not only due to the possible realization of a long-sought spin-liquid state and exotic Majorana excitations~\cite{Kitaev2006}, but also since in these materials anisotropic terms of the exchange interaction do manifest themselves in various physical observables such as magnetization~\cite{kubota2015,Janssen2016,Sebastian19_field_anisotropy,Maksimov2022}, thermal conductivity~\cite{Kasahara2018,Moore_swt_2018,Hess2020}, Raman spectra~\cite{sandilands2015,wulferding2020,sahasrabudhe2020} and spin-spin correlations \cite{Coldea12,us_rucl3,banerjee2018,Gog_Na2IrO3_2020}.

Another important factor affecting magnetic properties is dimensionality. It is well known that the temperature of a long-range magnetic order can be strongly reduced going from 3D to 2D or 1D \cite{MerminWagner,Katanin2007}. One-dimensional magnetic systems represent a limiting and very specific case, where each spin is connected with only two neighbors and no interchain exchange pathways to assist a long-range magnetic order are possible. This leads to a number of unexpected phenomena, including complete suppression of the ordered state, and dependence of the spin excitation spectra on spin-parity with the low-lying excitations very different from standard spin-waves~\cite{Haldane,faddeev}.

It was recently realized that these two intriguing concepts of Kitaev physics and one-dimensionality naturally meet in the case of CoNb$_2$O$_6$~\cite{Morris2021}. This material was considered as a model system to study quantum criticality in Ising chains~\cite{coldea2010,lee2010,Kinross2014,Sid_Co_2020}. The ground state of Co$^{2+}$ can be described by the effective total momentum $j_{\text{eff}}=1/2$ as in many other cobaltites, which are under active investigation now~\cite{Zhong2020,Coldea_Co_2020,Winter_Co_2022,Maksimov2022Ba,Vavilova2023}. Moreover, CoO$_6$ octahedra form zigzag chains sharing their edges. 
This particular geometry with 90$^{\circ}$ Co-O-Co bond can lead to the bond-dependent anisotropy of exchange interactions~\cite{Liu2018,Sano2018}. Indeed, Ising bond-dependent anisotropy (so-called ``twisted Kitaev chain'' model) was successfully applied to explain THz spectroscopy measurements~\cite{Morris2021}, although the angle between local Ising axes was found to be 34$^\circ$ (rather than the 90$^\circ$ for the ideal Kitaev chain), implying significant uniform Ising anisotropy. As such, an alternative related model, consisting of uniform Ising anisotropy and off-diagonal exchange, has also been used in the literature to describe results of inelastic neutron scattering (INS) experiments~\cite{Sid_Co_2020,woodland2023}. The bond-independent anisotropic exchange arises from the effects of local crystal field distortions on the composition of the $j_{\rm eff} = 1/2$ moments\cite{Winter_Co_2022}, making evaluation of the crystal field terms an important diagnostic of Co$^{2+}$ systems.

In this work, we establish the H-T phase diagram by thorough thermodynamic measurements and present the results of INS experiments analyzed by linear spin-wave theory (LSWT) and density functional theory (DFT) calculations of another material consisting of Co zigzag chains - pyroxene \scgo. 

Pyroxenes constitute a large class of natural minerals, forming $\sim$20 vol\% of the Earth's crust~\cite{deer1997,ringwood1994}. 
There are many pyroxenes with magnetic transition metals such as, e.g., aegirine NaFeSi$_2$O$_6$, chinese jade Na(Al,Fe)Si$_2$O$_6$, and kosmochlor NaCrSi$_2$O$_6$. They were shown to exhibit a variety of intriguing physical properties, such as multiferroicity~\cite{jodlauk2007}, orbital-Peierls~\cite{Wezel2006,Streltsov2006,Bozin2019} etc. and generally, they can be considered as quasi-1D magnets. In pyroxenes, the transition metals are octahedrally coordinated with oxygens forming zigzag chains separated by SiO$_4$ or GeO$_4$ tetrahedra as shown in Fig.~\ref{fig_crystallography}(a). 

\scgo\ orders magnetically below $T_\mathrm{N}=9$~K with spins coupled ferromagnetically in the chains and antiferromagnetic (AFM) order between chains, see Fig.~\ref{fig_crystallography}(a)~\cite{ding2016}. 
The effective magnetic moment in Curie-Weiss theory is larger than the spin-only value, which points to a significant orbital contribution. Together with strong magnetic anisotropy \cite{ding2016}, these measurements highlight the importance of spin-orbit physics in SrCoGe$_2$O$_6$.

\section*{RESULTS}
\subsection*{Thermodynamic measurements}
\begin{figure}
    \centering
    \includegraphics[width=0.9\columnwidth]{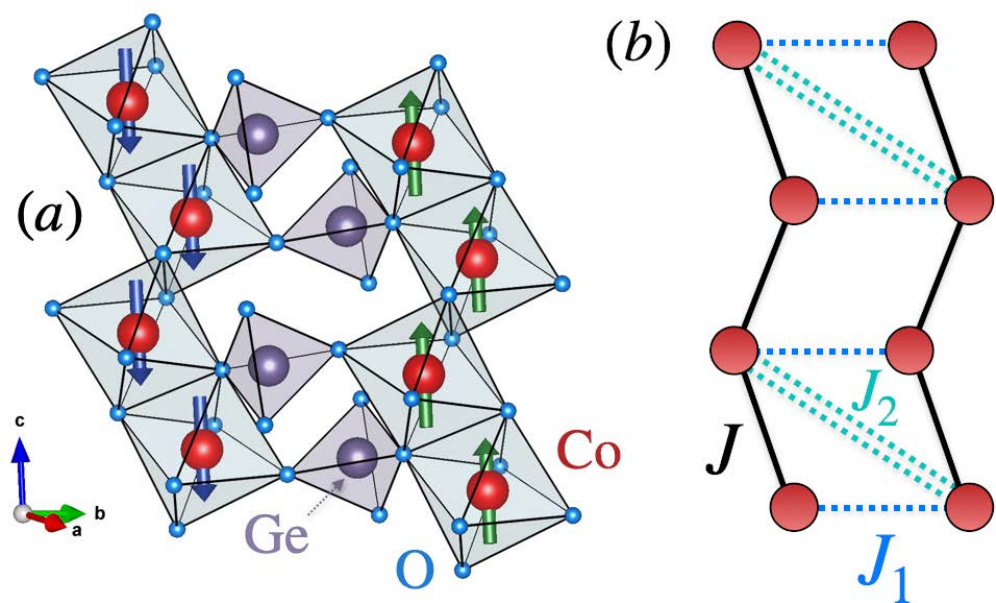}
    \caption{{\bf Structural details and exchange couplings in \scgo.} (a) Crystal structure and experimentally observed magnetic order in \scgo.  Strontium is not shown for clarity. Two different types of common edges between CoO$_6$ octahedra are shown by red and green lines. (b) A sketch of the spin sub-lattice: $J$ is intra-chain, $J_1$ and $J_2$ are inter-chain exchange parameters. While there are four adjacent chains to each Co-Co chain, only one is shown for clarity. All neighboring chains and the coordinate system are presented in the Methods section. }
    \label{fig_crystallography}
\end{figure}

\begin{figure}
\centering
    \includegraphics[width=0.9\columnwidth]{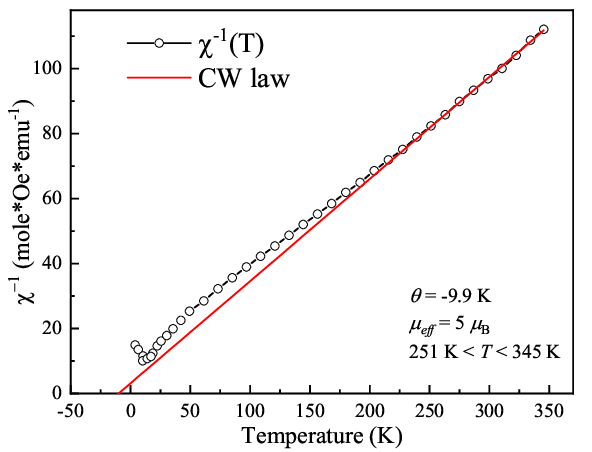}
    \caption{{\bf Curie-Weiss fit of magnetic susceptibility.} Inverse magnetic susceptibility $\chi_M(T)$ of polycrystalline sample of \scgo. High-temperature data is fit using Curie-Weiss law in \eqref{eq_CW}.}
    \label{fig_CW}
\end{figure}

\begin{figure*}[t!]
    \centering
    \includegraphics[width=1.9\columnwidth]{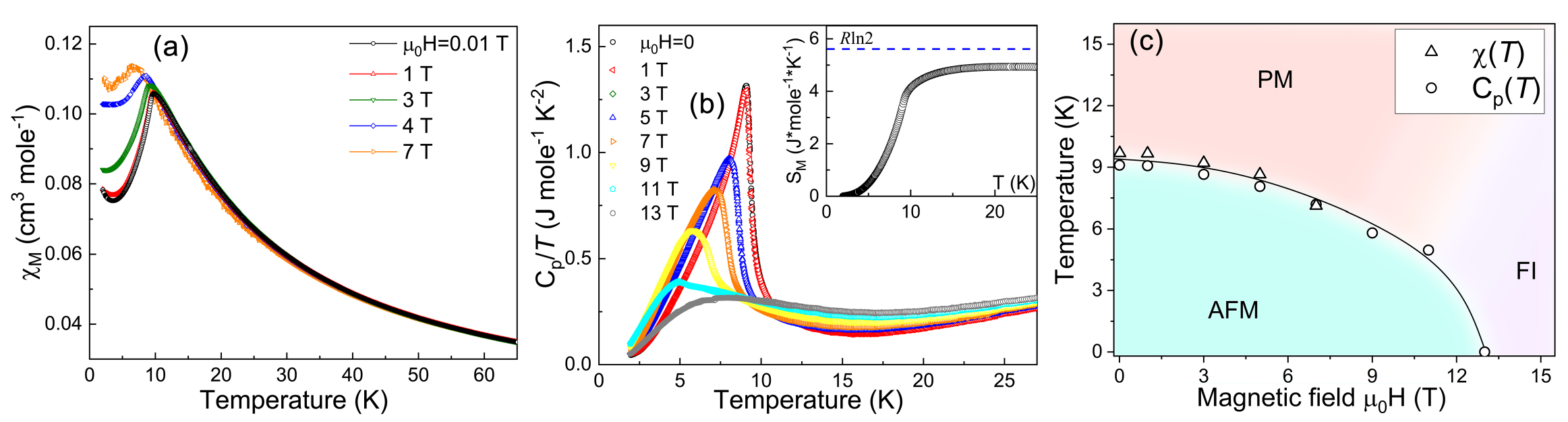}
    \caption{{\bf Magnetic susceptibility, heat capacity and magnetic phase diagram of \scgo.}
    (a) Magnetic susceptibility curves $\chi(T)$ of a polycrystalline sample of \scgo\ measured in applied magnetic fields after zero-field cooling (ZFC) procedure. (b) Heat capacity curves $C_p(T)/T$  of \scgo\ measured in applied magnetic fields up to $\mu_0H=13$~T. (c) Magnetic phase diagram of \scgo\ plotted in $T-\mathrm{\mu_0H}$ coordinates. PM, FI, and AFM mark paramagnetic state, field-induced state, and antiferromagnetic state, respectively.}
    \label{chi-c-PD}
\end{figure*}

We start with magnetic susceptibility data $\chi_M(T)$ measured on a polycrystalline sample of SrCoGe$_2$O$_6$ in applied magnetic fields up to $\mu_0H=7$~T after zero-field cooling (ZFC) procedure (see Figs.~\ref{fig_CW} and \ref{chi-c-PD}(a)). All the measured ZFC-curves reveal AFM-like behavior and well-defined anomaly around the N\'{e}el temperature similar to the susceptibility data reported in Ref.~\cite{ding2016}. Zero-field magnetic susceptibility can be fit using Curie-Weiss law:
\begin{align}
\chi_M(T)=\frac{C}{T-\theta_\text{CW}}.
\label{eq_CW}
\end{align}
We obtain $\mu_\text{eff}=5.0\mu_B$ and $\theta_\text{CW}=-9.9$ K by fitting high-temperature data in the range of 251~K~$<T<$~345~K. This value of $\theta_\text{CW}$ is in agreement with the AFM ground state (cf. Ref.~\cite{ding2016}).

The application of an external magnetic field partially suppresses the low-temperature AFM state and shifts the anomaly towards lower temperatures. 
No sign of saturation was observed on the isothermal magnetization curves $M(H)$ measured at $T=2$~K in magnetic fields up to 7~T (see SI, Supplementary Note 1 for details).

Similar behavior can be seen in the heat capacity data measured in applied magnetic fields up to $\mu_0H=13$~T and plotted as a function of temperature  $C_p(T)/T$, Fig.~\ref{chi-c-PD}(b).  The N\'{e}el temperature of SrCoGe$_2$O$_6$ was found to be $T_{\mathrm{N}}=9.1$~K as a peak of $\lambda$-type anomaly in zero magnetic field. On the contrary, no sign of the $\lambda$-type anomaly is observed on the $C_p(T)/T$ curve measured in the applied magnetic field $\mu_0H=13$~T implying complete suppression of the low-temperature AFM state in SrCoGe$_2$O$_6$. 
The field-temperature phase diagram, deduced from the heat capacity and magnetic susceptibility data, is summarized in Fig.~\ref{chi-c-PD}(c).

Another important piece of information that thermodynamic measurements provide is the degeneracy of the ground state. Temperature dependence of magnetic entropy was estimated after subtraction of the lattice contribution from our heat capacity data (see inset in Fig.~\ref{chi-c-PD}(b)). The magnetic entropy saturates close to $R\mathrm{ln}2$ value that is expected for a two-level system.  This clearly shows that the ground state is a doublet and suggests that the low-lying magnetic excitations can be described by $j_{\text{eff}}=1/2$ due to a sizable SOC despite strong distortions of CoO$_6$ octahedra.

\subsection*{Inelastic neutron scattering}

\begin{figure}[t!]
    \centering
    \includegraphics[width=1.0\columnwidth]{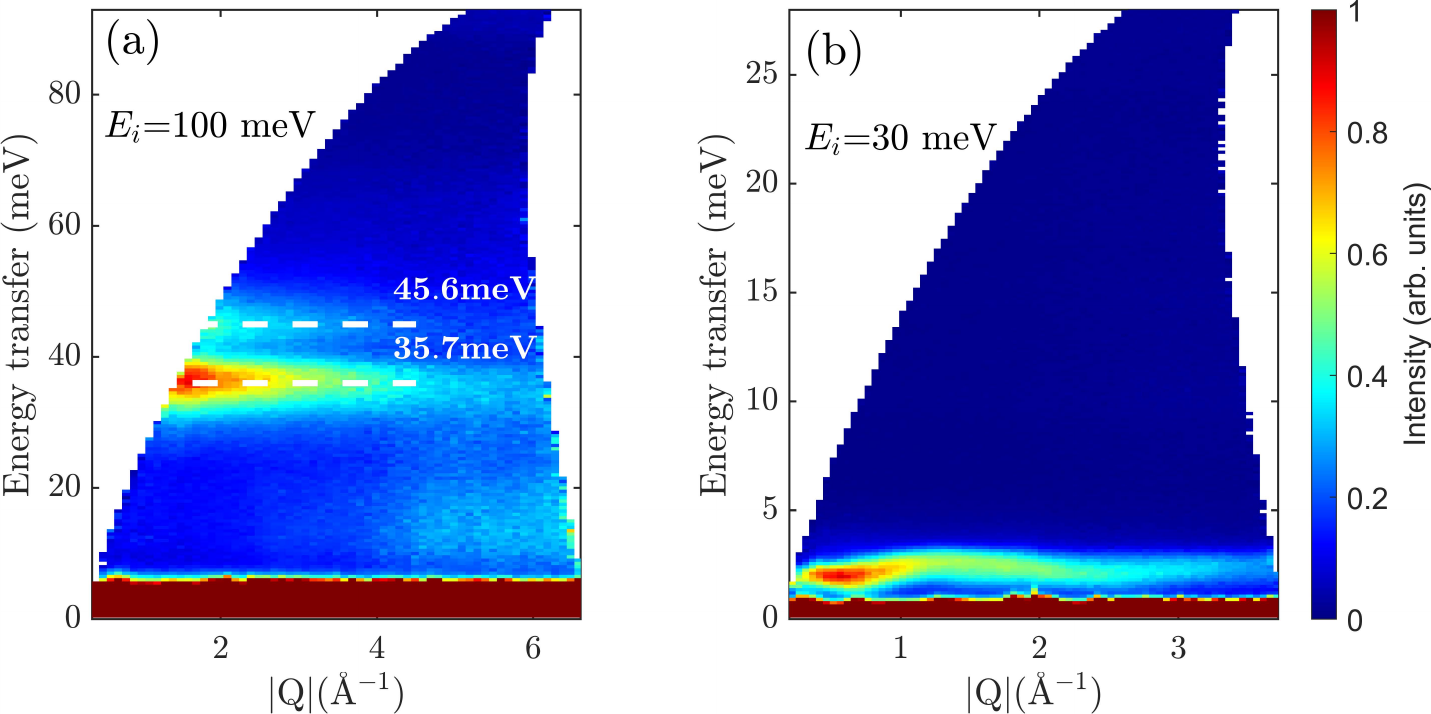}
    \caption{{\bf High-energy inelastic neutron scattering results.} INS taken at the SEQUOIA spectrometer at $T=6$~K with incident neutron energies $E_i=100$~meV (a) and $E_i=30$~meV (b), displaying CEF excitations in \scgo.}
    \label{cef}
\end{figure}

To demonstrate that the $j_{\text{eff}}=1/2$ doublet corresponding to the ground state of \scgo\ is well isolated from other states, first, we conducted high-energy INS measurements using incident energies $E_i=30$, 100, 150 and 250~meV (see SI, Supplementary Note 2 for the details of high-energy INS spectra).
The INS spectrum, depicted in Fig.~\ref{cef}(a), clearly exhibits two non-dispersive excitations at energy transfers of 35.7(3)~meV and 45.6(3)~meV. The intensity of these excitations decreases with increasing momentum transfer, providing definitive evidence that they correspond to ground-state crystal electric field (CEF) transitions of the Co$^{2+}$ ion.
Importantly, we did not observe any CEF levels below 30~meV, as illustrated in Fig.~\ref{cef}(b). 
We performed point charge model (PCM) calculations which show that the the ground state is a doublet and it is well separated from first two excited doublets (by $\sim$32 meV and $\sim$47 meV) in agreement with experimental data, for details see SI, Supplementary Note 5.
This finding, along with the complementary heat capacity and susceptibility data presented in Fig.~\ref{chi-c-PD}, supports the conclusion that the low-temperature magnetic properties of \scgo\ are primarily governed by the ground-state doublet, which effectively behaves as a $j_{\text{eff}}=1/2$ pseudospin. The observed CEF levels represent $j_{\rm eff} = 1/2 \to 3/2$ excitations, which are split by the distortion of the local crystal field away from ideal octahedral symmetry. Interestingly, in the INS study of another cobaltite CoNb$_2$O$_6$ with Co$^{2+}$ ions a very similar level structure was observed with excitation energies $\sim$ 30 and 50 meV and it was successfully explained by a model based on pseudospin $j_\text{eff}=1/2$ \cite{ringler2022}. However, it is worth noting that the splitting of the $j_{\rm eff} = 3/2$ levels is significantly smaller in \scgo, which points to a potentially reduced effect of bond-independent anisotropies in the magnetic couplings.

Figure~\ref{cef}(b) provides clear evidence of a dispersive spin-wave excitation extending up to approximately 3~meV, indicating significant couplings in \scgo. 
However, the spectra obtained at high incident energies lack the necessary energy resolution to distinguish or resolve these low-energy features adequately. 
To further investigate the spin dynamics of the Co magnetic sublattice, we conducted measurements on the Cold Neutron Chopper Spectrometer (CNCS) with $E_i=5.0$~meV at a temperature of 1.7~K.

\begin{figure*}[t!]
    \centering
    \includegraphics[width=2.0\columnwidth]{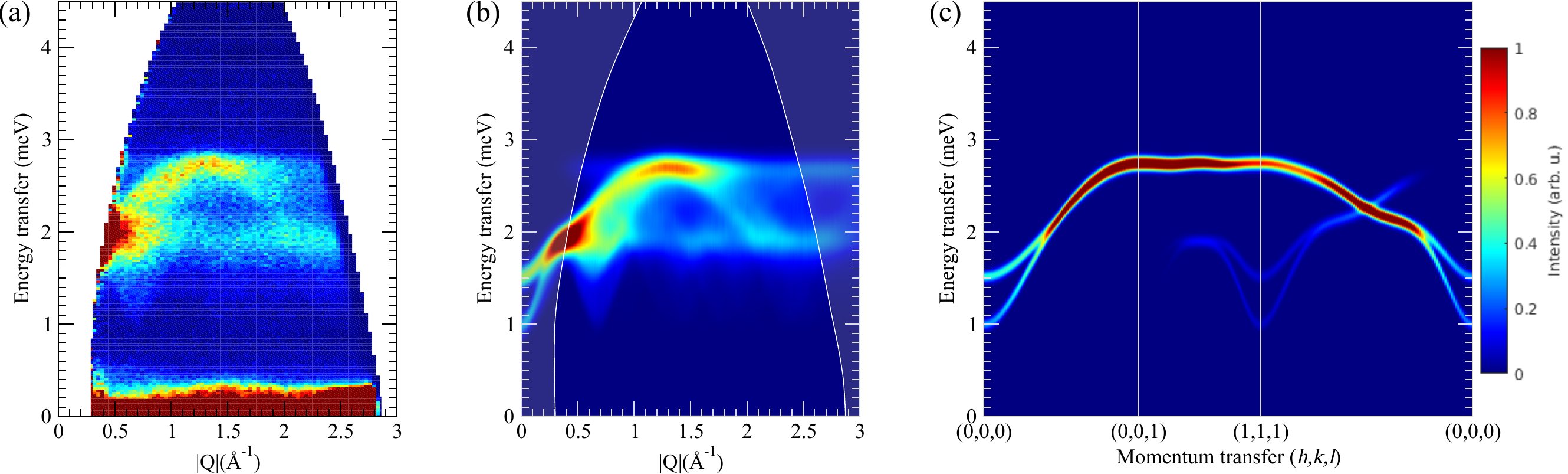}
    \caption{{\bf Comparison between the experimental and calculated inelastic neutron spectra in \scgo.} (a) INS data taken at the CNCS spectrometer with incident neutron energy $E_i=5.0$~meV. (b) Linear spin-wave calculation for the powder averaged spectrum for the best-fit set of parameters, see text. (c) Same, without averaging, along the contour involving high-symmetry points.}
    \label{fig_powder}
\end{figure*}

The experimentally observed low-energy INS as a function of energy and momentum transfer is shown in Fig.~\ref{fig_powder}(a).
The excitation spectrum is dominated by a strong dispersive magnon-like mode within the 2 to 3~meV energy range. 
A second excitation at around 2~meV appears as a dispersionless band with modulated intensity across the entire Q region.
An evident concave shape, exhibiting a minimum around $|\mathrm{Q}|\sim 0.7$~\AA$^{-1}$, is observable in the data. 
This characteristic is reminiscent of the behavior observed in other honeycomb lattice magnets with zigzag AFM order~\cite{banerjee2016, Ma_Co_2020}.
No magnetic signals were observed above an energy transfer of 3~meV.\\

Note, that the diffraction patterns obtained by integrating elastic scattering in the range [-0.2, 0.2]~meV at temperatures 1.7~K and 20~K support the magnetic transition and agree well with the reported data \cite{ding2016}, see SI, Supplementary Note 3.

\subsection*{First principle calculations}

Careful analysis of the INS can be performed within the spin-wave theory, but to reduce the number of model parameters and also narrow their value range we first perform DFT calculations.

SrCoGe$_2$O$_6$ was found to be an insulator with the band gap of $1.8$~eV and magnetic moments $2.8$~$\mu_B$ in GGA+U calculations for Hubbard $U$ = 6 eV and Hund's $J_H$ = 0.9 eV, typically used in literature~\cite{Maksimov2022Ba}. Account of the SOC via GGA+U+SOC shows that spins are predominantly oriented along $c$ axis and orbital contribution to the magnetic moment is $0.2$~$\mu_B$. 

The spin lattice in pyroxenes can be characterized by chains connected by two inter-chain couplings $J_1$ and $J_2$, see  Fig.~\ref{fig_crystallography}(b) and Methods section. These inter-chain exchanges are mediated via hopping through GeO$_4$ tetrahedra. Direct calculation of the exchange coupling by the four-state method~\cite{Li2021} reveals that for Hamiltonian defined  in \eqref{eq_H_J1_J2} inter-chain exchanges are $J_1 = 0.74$~meV and $J_2 = 1.06$~meV (AFM). For intra-chain coupling we estimated both isotropic component $J = -1.20$~meV (FM) and Kitaev exchange $K = 1.12$~meV. While these results can not be considered as a ``final truth'', since they do not take into account a multiplet structure of excited states (which can be important for cobaltites~\cite{Liu2023,Winter_Co_2022}) they point to importance of both Kitaev interaction and inter-chain coupling. In fact large $J_1$ and $J_2$ does not seem surprising, since Ge $4p$ orbitals are known to be rather extended and can substantially increase this interaction~\cite{Streltsov2008}. Strong inter-chain exchange implies that long-range order is stable towards fluctuations and spin-wave theory can be used to describe magnetic excitations, instead of spinons such as the case of CoNb$_2$O$_6$ \cite{Sid_Co_2020}.

\subsection*{Spin-wave theory}

The gapped nature of the magnetic excitation spectrum, as shown in Fig.~\ref{fig_powder}(a), implies that the isotropic Heisenberg model is not enough to describe its properties. Moreover, the spin-orbit entangled structure of the $j_\text{eff}=1/2$ pseudospin allows for anisotropic interactions, given they obey the symmetry of the lattice.

For SrCoGe$_2$O$_6$, similarly to previously studied CoNb$_2$O$_6$ \cite{Sid_Co_2020}, the glide symmetry of the zigzag chains and bond inversion symmetry allows for six intra-chain nearest-neighbor couplings:
\begin{align}
\label{eq_Hmatrix}
J^{(m)}=\begin{pmatrix}
J^{xx} & (-1)^m J^{xy} & J^{xz}\\
(-1)^m J^{xy} & J^{yy} & (-1)^m J^{yz}\\
J^{xz} & (-1)^m J^{yz} & J^{zz}
\end{pmatrix},
\end{align}
where $m=1,2$ corresponds to two types of bonds in the staggered structure of the chains. The corresponding reference frame for spin operators and two types of bonds, denoted by $x$ and $y$, are shown in the Methods section.

Moreover, DFT calculations predict two large inter-chain interactions. Thus, there are eight parameters in the model even without taking into account the anisotropy of $J_1$ and $J_2$. With rather limited information on the powder INS data, the spin-wave fit of the full model becomes extremely challenging.
Therefore, it is advisable to consider a simplified model with fewer free parameters, which can still sufficiently fit the inelastic powder data.

We argue that the ground state and spin-wave spectrum of SrCoGe$_2$O$_6$ can be described by two inter-chain exchanges and the one-dimensional version of the extended Kitaev-Heisenberg (KH) model for intra-chain interactions:
\begin{align}
\mathcal{H}=\mathcal{H}_\text{KH}+\sum_{\langle ij \rangle_1}
J_1 \mathbf{S}_i \cdot \mathbf{S}_j+\sum_{\langle ij \rangle_2}
J_2 \mathbf{S}_i \cdot \mathbf{S}_j,
\label{eq_H_J1_J2}
\end{align}
where exchanges between ions in the chain are given by
\begin{align}
\mathcal{H}_\text{KH}=\sum_{\langle ij \rangle_\gamma}&
J \mathbf{S}_i \cdot \mathbf{S}_j +K S^\gamma_i S^\gamma_j 
+\Gamma \left( S^\alpha_i S^\beta_j +S^\beta_i S^\alpha_j\right)\nonumber\\
+&\Gamma' \left( S^\gamma_i S^\alpha_j+S^\gamma_i S^\beta_j+S^\alpha_i S^\gamma_j+S^\beta_i S^\gamma_j\right).
\label{eq_H_JKGGp}
\end{align}
Here $\langle ij \rangle_{1,2}$ stand for two types of inter-chain bonds, see Fig.~\ref{fig_crystallography}, while $\{\alpha,\beta,\gamma\}=$\{y,z,x\} at the $x$-bond of the chain, and \{z,x,y\} at the $y$-bond. The reasoning for this choice of the intra-chain exchange model is the fact that the edge-sharing octahedra and staggered zigzag chains can be viewed as a honeycomb model with one leg missing \cite{1d_Kitaev}, also referred to as Kitaev spin chain \cite{Oles2007,Affleck_PRR_2020,Morris2021,Gallegos}. Therefore, we use a four-parameter model for Co chains as an ansatz, which obeys the symmetries of the general Hamiltonian \eqref{eq_Hmatrix}, and represents a subset of the full model with parameters being related as
\begin{align}
\label{eq_fulltoK}
J_{xx}&=J+\frac{1}{6}\left(K-4\Gamma-2\Gamma'\right),\\
J_{yy}&=J+K/2-\Gamma',\nonumber\\
J_{zz}&=J+\frac{1}{3}\left( K+2\Gamma+4\Gamma'\right),\nonumber\\
J_{xy}&=\frac{1}{2\sqrt{3}}\left(K+2\Gamma-2\Gamma'\right),\nonumber\\
J_{xz}&=\frac{1}{3\sqrt{2}}\left(-K+\Gamma-\Gamma'\right),\nonumber\\
J_{yz}&=-\frac{1}{\sqrt{6}}\left(K-\Gamma+\Gamma'\right).\nonumber
\end{align}
While we have to acknowledge that it is the full six-parameter model that needs to be used to describe intra-chain interactions of SrCoGe$_2$O$_6$, some properties are naturally described by the extended Kitaev-Heisenberg model. As we show in the Methods section, there is a parameter regime of the intra-chain ferromagnetic state with magnetic moments in the $x$-$z$ plane, which is the state observed experimentally \cite{ding2016}. We refer to this state as FM-$xz$, it is illustrated in the Methods section. Thus, fixing spin direction can immediately place constraints on the parameters of the Kitaev-Heisenberg model \cite{chaloupka2015}:
\begin{align}
    \tan 2\theta=4\sqrt{2}\frac{1+r}{7r-2},~r=-\frac{\Gamma}{K+\Gamma'},
    \label{eq_angle}
\end{align}
where $\theta\approx6^\circ$, according to measurements on SrCoGe$_2$O$_6$ \cite{ding2016}. However, the anisotropy of $g$-factors, which is unknown, can modify this value. Thus, to provide a reasonable fitting to neutron-scattering data we limit our model for SrCoGe$_2$O$_6$ to four intra-chain exchanges and two inter-chain exchanges.

We use LSWT to calculate the magnetic spectrum in the AFM state where chains are ordered ferromagnetically in the FM-$xz$ configuration. Due to the scrambled nature of the neutron scattering on the powder sample, the spin-wave fit may be challenging, especially due to the large parameter phase space. However, there are three discernible features of the spectrum that we can use for the fits of magnon energies, as one can infer from Fig.~\ref{fig_powder}(a): (i) the minimum of intensity at 1 meV and $|\mathbf{Q}|\simeq 0.7\text{~\AA}^{-1}$; (ii) the flat band at 1.9 meV; (iii) a well-defined maximum at 2.7 meV. By studying spin-wave spectra for various parameter sets we found that the first feature originates in $\mathbf{k}=0$ mode, as one can see in Fig.~\ref{fig_powder}(c), while 1.9 and 2.7 meV features can be traced to $(1/2,1/2,0)$ modes. This simplifies calculations significantly since these magnon energies can be obtained analytically, see Methods section.

\begin{figure}[t!]
    \centering
    \includegraphics[width=0.99\columnwidth]{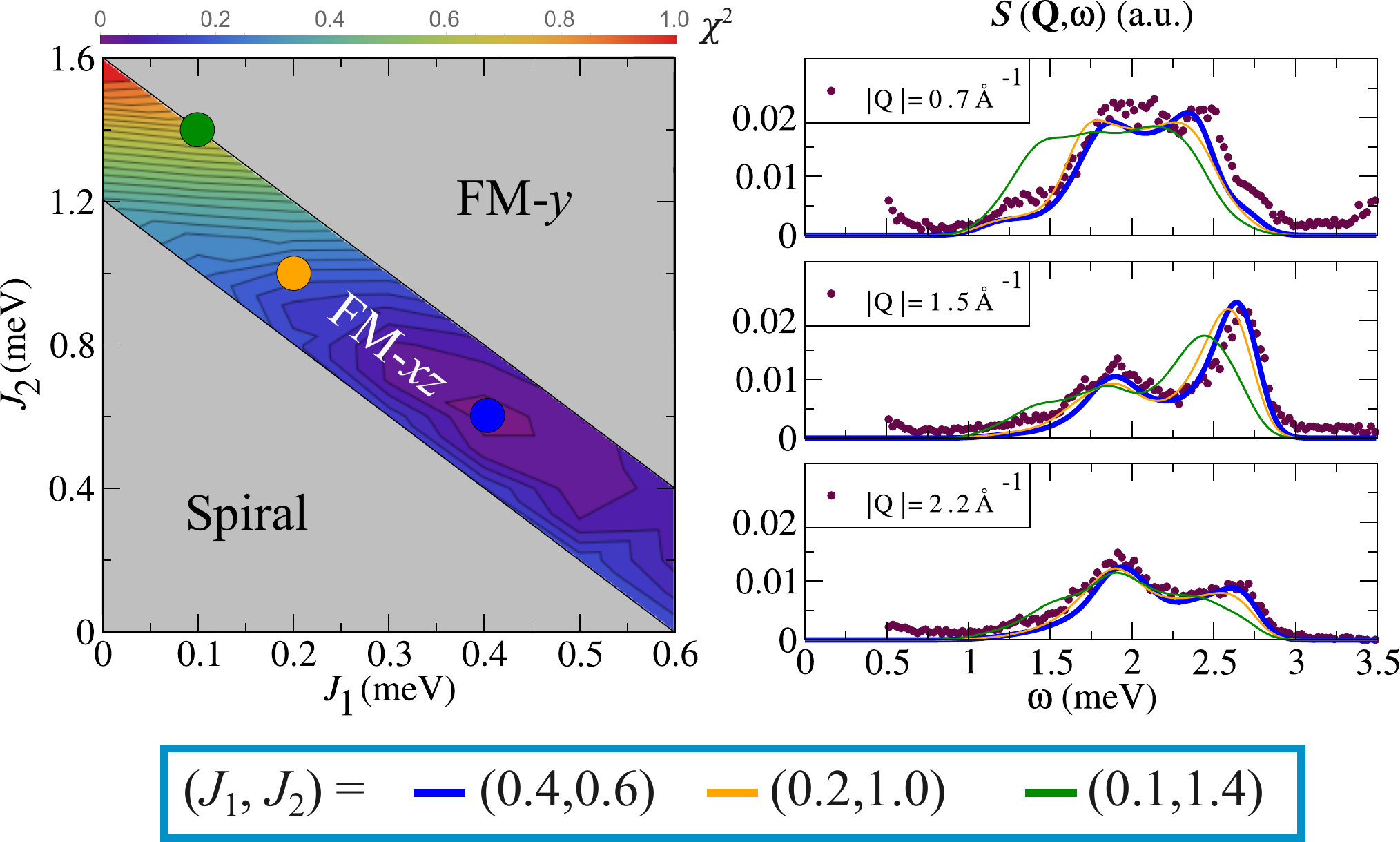}
    \caption{{\bf Determination of exchange interactions in SrCoGe$_2$O$_6$ from neutron scattering data.} (left) Fit quality $\chi^2$, shown as a function of the inter-chain parameters $J_1$ and $J_2$ [Fig. 1(b)], see text. The blue dot indicates the position of the global minimum. (right) Constant-Q cuts through the experimental (symbols) and calculated (solid lines) spectral functions for three representative Q values.}
    \label{fig_fit}
\end{figure}

Thus, we can use a combination of three criteria for magnon energies from neutron-scattering data and the tilt angle \eqref{eq_angle} to use for the analytical fitting of four criteria of the intra-chain exchange matrix. We keep inter-chain exchanges as free parameters, and for each set of $(J_1,J_2)$ we obtain $(J,K,\Gamma,\Gamma')$ from fitting analytical expressions to the four criteria above. Next, we scan through the $J_1 - J_2$ phase space and obtain the dynamical structure factor using SpinW package \cite{tothlake} to compare to the CNCS neutron-scattering data. During the calculation, a Gaussian broadening with $\sigma=0.2$ meV was applied. The result is shown in Fig.~\ref{fig_fit} as an intensity plot of fit quality
\begin{align}
\chi^2=\sum_{Q,\omega}\left( \mathcal{S}^\text{exp}(Q,\omega)-\mathcal{S}^\text{calc}(Q,\omega)\right)^2.
\end{align}
Note that for some parameter sets the ground state is not FM-$xz$ relevant to \scgo, according to classical minimization but instead spiral or FM-$y$ (shown in Fig.~\ref{fig_chain_pd}) state, which we also use to limit parameters applicable to \scgo. According to the definition above the best fit to the powder scattering data is given by the minimum of the fit quality $\chi^2$ in Fig.~\ref{fig_fit}, where $\chi^2$ is normalized to the minimum and maximum values. The best-fit parameter set is obtained with  $J_1=0.4$ meV and $J_2=0.6$ meV. The optimal intra-chain exchanges for these values of inter-chain interactions are given by \begin{align}
J=-0.87,~K=0.83,~\Gamma=0.43,~\Gamma'=-0.26\text{~(meV)}
\label{eq_parameters}
\end{align}
The spin-wave calculation for this set of parameters is shown in Fig.~\ref{fig_powder} with powder averaged signal shown in (b), and the dynamical structure factor along a contour in the BZ in (c). One can see that this parameter set provides an excellent agreement with CNCS data in Fig.~\ref{fig_powder}(a). As an illustration of the quality of the fit, in Fig.~\ref{fig_fit}, right panel, we plot one-dimensional cuts of the spin-spin correlation function for several values of $|\mathbf{Q}|$, together with experimental data. We plot these cuts for three sets of parameters, the optimal ones (blue lines) and two more, shown with orange and green lines. The optimal parameter set matches neutron-scattering data excellently, except for some discrepancy at low energies and small $|\mathbf{Q}|$. Moreover, this parameter set yields $\theta_\text{CW}=-4.67$~K in a solid agreement with the susceptibility fit in Fig.~\ref{fig_CW}.

\section*{DISCUSSION}

The exchange parameters presented in \eqref{eq_parameters} were obtained by fitting of the powder inelastic neutron spectra data, all tendencies are fully consistent with results of ab initio calculations, see Tab.~\ref{LSWTvsDFT}. This strongly supports our main findings: (i) importance of Kitaev interaction, which is $K \sim |J|$ (this is in a sense similar to famous RuCl$_3$, where $K \sim \Gamma$ \cite{Winter_review}), and (ii) substantial inter-chain coupling.

Nevertheless, while Kitaev bond-dependent anisotropic exchange interaction naturally appears in the shared-edge geometry~\cite{Liu2018,Sano2018}, we would like to stress that the choice of the Kitaev-Heisenberg model as an ansatz for \scgo\ is not unique. 

Interestingly, the matrix of the Kitaev-Heisenberg model in the local quantization axes for the best-fit parameters \eqref{eq_parameters} has off-diagonal term $\mathrm{J}^{xy}$ (the only off-diagonal term contributing linear spin-wave spectrum) smaller than diagonal terms:
\begin{align}
\hat{\mathrm{J}}^{\text{loc}}=\begin{pmatrix}
-0.65 & \mp 0.13 & 0\\
\mp 0.13 & -0.19 & \pm 0.63\\
0 & \pm 0.63 & -0.93
\end{pmatrix} \text{(meV).}
\end{align}

The fact that off-diagonal terms $\mathrm{J}^{xy}$ are smaller than diagonal ones motivated us to study an alternative model where we neglect small off-diagonal terms and come to an XYZ Hamiltonian in the local coordinate systems with three independent parameters: $\mathrm{J}^{xx}$, $\mathrm{J}^{yy}$, and $\mathrm{J}^{zz}$. In this case, the direction of spins is determined, not by anisotropic exchanges but by the distortions of the lattice and subsequent rotation of quantization axes \cite{Winter_Co_2022}. We should note that, if available, INS measurements on single crystal samples would allow one to extract all six exchanges of the matrix \eqref{eq_Hmatrix} and test the viability of Kitaev-Heisenberg model for \scgo.

We have performed spin-wave calculations for this model and the best-fit powder spectrum provides a satisfactory agreement with neutron scattering data, albeit still slightly worse than the Kitaev-Heisenberg model. This fact implies that the exchange matrix for \scgo~ is most likely close to the diagonal-only model but off-diagonal terms $\mathrm{J}^{xy}$ are non-negligible and improve the fitting of neutron-scattering data. Moreover, smaller splitting of CEF levels in \scgo, as shown in Fig.~\ref{cef}, compared to CoNb$_2$O$_6$\cite{ringler2022}, points to proximity to the cubic limit and the relevance of Kitaev-Heisenberg interactions in \scgo, instead of Ising model. 

In summary, we successfully synthesized powder samples of SrCoGe$_2$O$_6$, and performed a detailed study of magnetic excitations by the inelastic neutron scattering, which were analyzed by a combination of \textit{ab initio} and spin-wave theory calculations. These results demonstrate substantial bond-dependent Kitaev exchange, $K/|J|=0.96$, while our heat capacity measurements revealed that the external magnetic field of $\sim$ 13~T transforms a fragile antiferromagnetic ordering with $T_c \approx 9$~K to a field-induced state widely discussed in context of Kitaev physics in other cobaltites~\cite{Songvilay_Co_2020,Park_Co_2020,NCTO_Garlea_2021,Rachel_Co_2021}. Thus, our study demonstrated that pyroxenes can be considered as a new platform for the Kitaev model and, moreover, these materials can be made more one-dimensional changing Ge by Si with less extended $p$ orbitals. This can open up even more exciting perspectives for one-dimensional Kitaev-Heisenberg model, which exhibits exotic phases with fractionalized excitations \cite{Affleck_PRR_2020}, as well as Kitaev materials in general. 

\begin{table}[t!]
\centering
\caption{Comparison of the exchanges from ab initio (GGA+U+SOC) calculations and neutron scattering fit (LSWT). The exchanges are in units of meV. \label{LSWTvsDFT}}
\begin{tabular}{lrrrrrrr}
Method & $J$ & $K$ &$|K/J|$ &$\Gamma$ & $\Gamma'$ & $J_1$ & $J_2$ \\
GGA+U+SOC  & -1.20 & 1.12 & 0.93 & -    & -     & 0.74  & 1.06 \\
LSWT & -0.87 & 0.83 & 0.96 & 0.43 & -0.26 & 0.40  & 0.60 \\
\end{tabular}

\end{table}

\subsection*{Sample synthesis}
Polycrystalline powder of \scgo\ was prepared as a 10~g batch from a stoichiometric mixture of SrCO$_3$, Co$_3$O$_4$ and GeO$_2$. Starting materials were carefully homogenized by grinding for 20 minutes under isopropanol in an agate mortar, pressed to pellets and heated in open platinum crucibles in a chamber furnace at a temperature of 1473~K over a period of 21 days with four intermediate regrinding until phase pure and a good crystalline sample was obtained. Phase purity was checked by powder x-ray diffraction using a PANalytical X’Pert Pro MPD equipped with an X’Celerator solid-state detector (for details, see Supporting Information SI.4). 

\subsection*{Magnetization and heat capacity}
For magnetization measurements, 24.6~mg of \scgo\ were loaded on a Quantum Design MPMS~3 magnetometer. The sample was contained in weight paper and affixed with Apiezon~M grease.
Temperature-dependent magnetization data were collected under zero-field cooling conditions and sweep mode at several fields (100~Oe and 1, 3, 5, 7~T) from 2~K to 100~K. 
Companion measurements were performed at 100~Oe and 7~T under settle between 2~K and 50~K every 1~K mode to check for temperature stability.
Magnetization under a magnetic field was collected on the same sample on a single quadrant, with increasing field up to 7~T and at several temperatures (2, 10 and 50~K).

Heat capacity was measured using a Quantum Design Physical Property Measurement System. The \scgo\ powder was mixed with silver powder in equal parts by mass and pressed into a thin pellet. The pellet was cut with a razor blade to produce a 6.08~mg sample that was mounted using Apiezon N-grease for the heat capacity measurement. The heat capacity of the \scgo\ sample was determined by subtraction of the measured heat capacity data for silver from the total.

\subsection*{Inelastic neutron scattering}
INS measurements were performed with the use of the Cold Neutron Chopper Spectrometer CNCS~\cite{CNCS1,CNCS2} and the Fine Resolution Chopper Spectrometer SEQUOIA~\cite{SEQ}  at the Spallation Neutron Source (SNS) at Oak Ridge National Laboratory.
A powder sample of \scgo\ was loaded in an aluminum can for the measurements and a standard orange cryostat was used to cover the temperature region from 1.7~K to 100~K at CNCS, and bottom-loading closed cycle refrigerator ($T=5.5$~K) at SEQUOIA.
Data were collected using fixed incident neutron energies of 250, 150, 100, and 30~meV for SEQUOIA, and 5.0~meV for CNCS.
In these configurations, a full width at half maximum (FWHM) resolution of 18, 10, 6.1, and 1.0~meV (SEQUOIA), and 0.15~meV (CNCS) was obtained at the elastic position.
For data reduction and analysis we used the \textsc{Mantid}~\cite{Mantid} and \textsc{Dave}~\cite{Dave} software packages.

\subsection*{DFT calculations}
The \textit{ab initio} band structure calculations of SrCoGe$_2$O$_6$ were carried out within the framework of density functional theory realized in VASP package~\cite{Kresse-96}. We used the generalized gradient approximation (GGA)~\cite{Perdew-96}  and projector augmented wave (PAW) method~\cite{Blochl-94}. The exchange-correlation functional in Perdew-Burker-Ernzerhof (PBE) form was applied~\cite{Perdew-97}. Spin-orbit coupling was taken into account in the calculations. The cutoff energy was chosen to be 600 eV and a grid of 4$\times$4$\times$4 points was used for integration over the Brillouin zone. The correlation effects were taken into account via GGA+U approach\cite{Liechtenstein-95}. We chose the onsite Coulomb repulsion parameter to be U = $6$ eV and Hund's rule coupling parameter was taken as J$_\mathrm{H}$ = $0.9$ eV as typically used in literature for Co ions. The occupation numbers of Co-$3d$ states were obtained by integration within an atomic sphere with a radius $1.302$~\AA. The exchange interaction parameters were calculated using the four-state method~\cite{Li2021}.

\subsection*{Model conventions and spin-wave theory}
In general, bilinear exchange Hamiltonian can be represented in a matrix form 
\begin{align}
\hat{\cal H}=\sum_{\langle ij\rangle} \mathbf{S}^{\rm T}_i \hat{\bm J}_{ij} \mathbf{S}_j,
\label{eq_Hij}
\end{align}
where $\hat{\bm J}_{ij}$ is a $3\times 3$ matrix and can depend on the site position, such as the matrix of the model \eqref{eq_Hmatrix} being different on $x$ and $y$ bonds. If the exchange matrix $\hat{\bm J}_{ij}$ is not isotropic Heisenberg interaction, then its elements become reference frame dependent.

\begin{figure}[tbh]
    \centering
    \includegraphics[width=1.0\columnwidth]{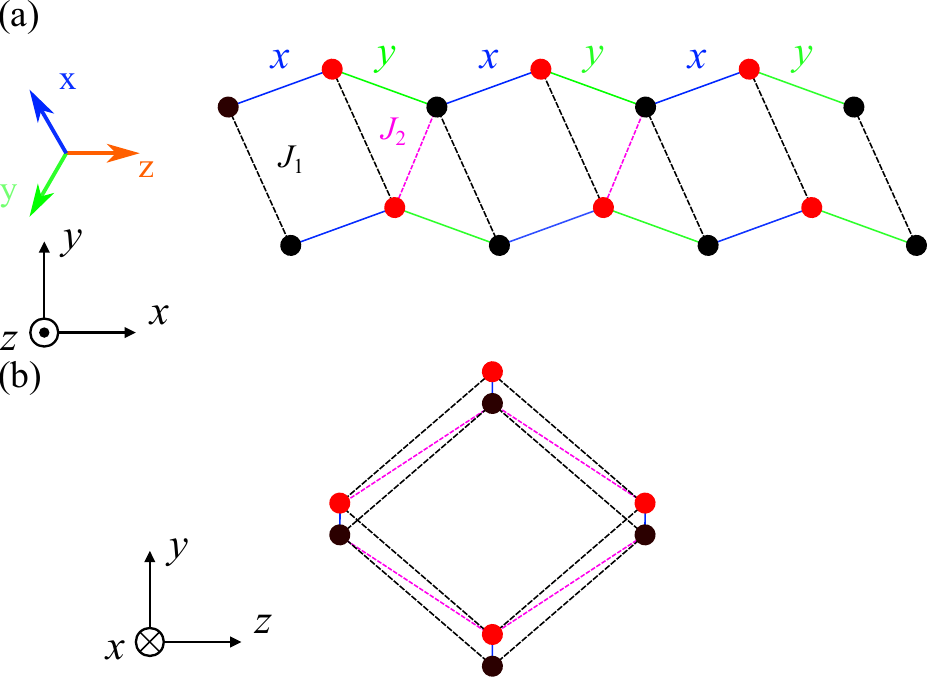}
    \caption{{\bf Schematic illustration of the exchange Hamiltonian \eqref{HJpm}.} Due to the staggered structure of Co-chains, there are two types of intra-chain bonds: $x$ and $y$. Two inter-chain exchanges are allowed due to the exchange paths through Ge ions.}
    \label{fig_exchanges}
\end{figure}

For instance, the extended Kitaev-Heisenberg model \eqref{eq_H_JKGGp} is written in cubic axes \{x,y,z\}, which are defined by the ideal ligand octahedra of cobalt-oxygen bonds, shown in Fig.~\ref{fig_exchanges}(a). However, it is also beneficial to represent this model in the crystallographic reference frame $\{x,y,z\}$, also shown in Fig.~\ref{fig_exchanges}(a). The transformation from cubic to crystallographic axes is given by $\hat{\bm J}^\text{cryst}_\alpha\!=\!\hat{\mathbf{R}_c^{-1}} \hat{\bm J}^\text{cubic}_\alpha \hat{\mathbf{R}}_c$, where  ${\bf S}_{\rm cubic}\!=\!
\hat{\mathbf{R}}_c{\bf S}_{\rm cryst}$, and
\begin{align}
\hat{\mathbf{R}}_c=\left(
\begin{array}{ccc}
 \frac{1}{\sqrt{6}} & -\frac{1}{\sqrt{2}} & \frac{1}{\sqrt{3}} \\
 \frac{1}{\sqrt{6}} &   \frac{1}{\sqrt{2}} & \frac{1}{\sqrt{3}} \\
 -\frac{2}{\sqrt{6}} & 0 & \frac{1}{\sqrt{3}} \\
\end{array} 
\right).
\label{eq_cubic_transform}
\end{align}

Using the transformation above, we yield the extended Kitaev-Heisenberg model in the crystallographic reference frame as
\begin{align}
{\cal H}_\text{cryst}=&\sum_{\langle ij\rangle}
J_\text{iso} \Big(S^{x}_i S^{x}_j+S^{y}_i S^{y}_j+\Delta S^{z}_i S^{z}_j\Big)\nonumber\\-&2 J_{\pm \pm} \Big( \Big( S^x_i S^x_j - S^y_i S^y_j \Big) c_\alpha 
-\Big( S^x_i S^y_j+S^y_i S^x_j\Big)s_\alpha \Big)\nonumber\\
 -&J_{z\pm}\Big( \Big( S^x_i S^z_j +S^z_i S^x_j \Big) c_\alpha 
 +\Big( S^y_i S^z_j+S^z_i S^y_j\Big)s_\alpha \Big),
\label{HJpm}
\end{align}
where $c_\alpha (s_\alpha)=\cos (\sin)\varphi_\alpha$, here bond-dependent phases are  $\varphi_x=2\pi/3$, $\varphi_y=-2\pi/3$. Note that it contains bond-isotropic $XXZ$ part and two-bond-dependent exchanges. These exchanges are related to the extended Kitaev-Heisenberg model via
\begin{align}
J_\text{iso}&=J+\frac{1}{3} \left( K-\Gamma-2\Gamma'\right),\nonumber\\
\Delta J_\text{iso}&=J+\frac{1}{3} \left( K+2\Gamma+4\Gamma'\right),\nonumber\\
2J_{\pm\pm}&=\frac{1}{3} \left( -K-2\Gamma+2\Gamma'\right),\nonumber\\
\sqrt{2}J_{z\pm}&=\frac{1}{3} \left( 2K-2\Gamma+2\Gamma'\right),
\label{eq_ice_to_K}
\end{align}

\begin{figure}[tbh]
    \centering
    \includegraphics[width=1.0\columnwidth]{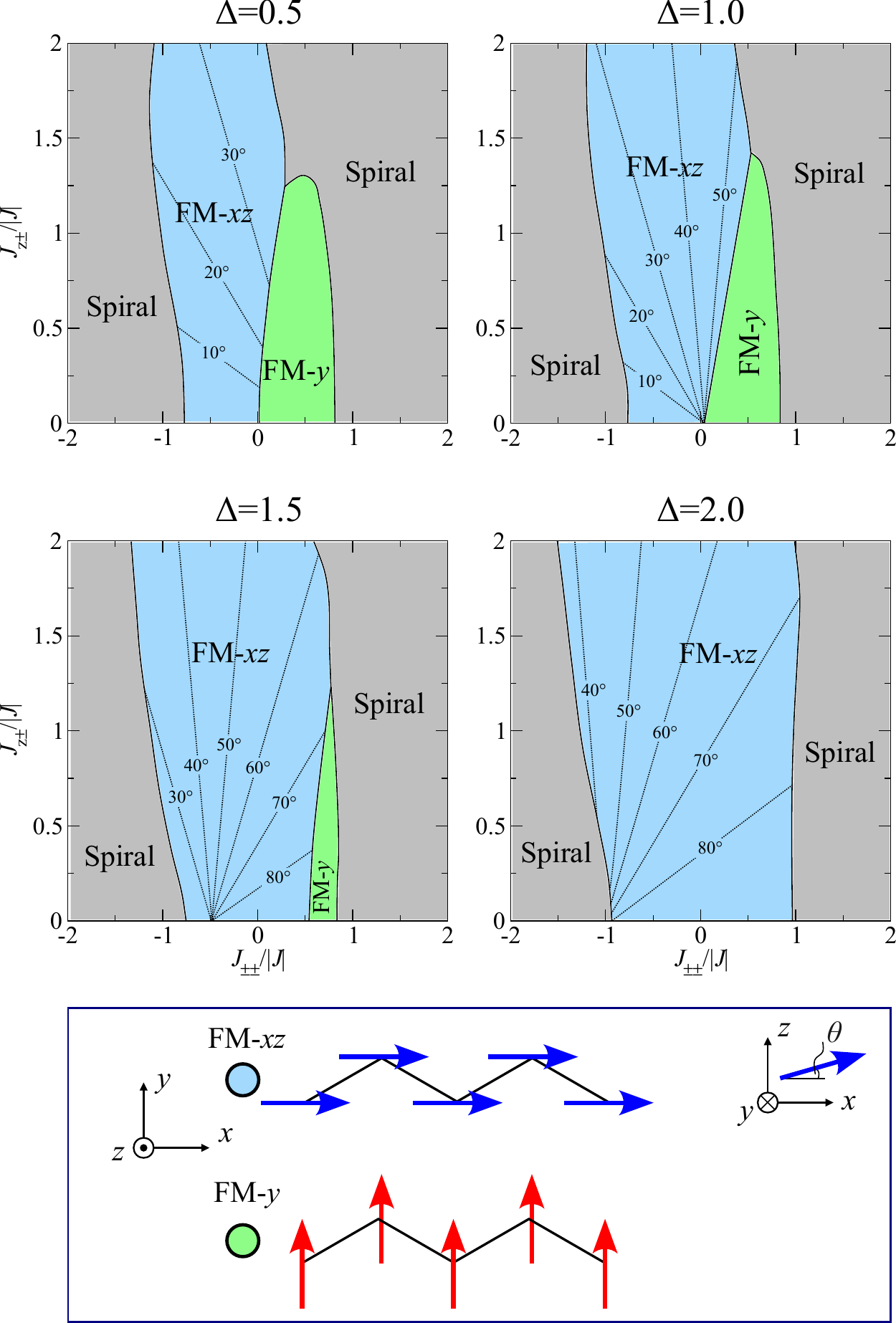}
    \caption{{\bf {Classical phase diagrams of the model \eqref{HJpm} for various XXZ anisotropy $\Delta$ with the structure of the ferromagnetic states}}. FM-$y$ is the ferromagnetic ordered state with magnetic moments along the $y$-axis, shown in Fig.~\ref{fig_exchanges}. FM-$xz$ state is a ferromagnetic ordered state where magnetic moments are in the $x$-$z$ plane. The angle of canting from the $x$-$y$ plane is shown on the phase diagrams.}
    \label{fig_chain_pd}
\end{figure}

Phase diagram of the model \eqref{HJpm}, obtained using Luttinger-Tisza method \cite{lt_original} is shown in Fig.~\ref{fig_chain_pd}. Note that inter-chain interactions are not included here. We plot slices of the phase space for $J_\text{iso}<0$ at fixed $\Delta$. For large $|J_{\pm\pm}|$ we observe incommensurate spiral states, while for smaller $|J_{\pm\pm}|$ there are two ferromagnetic states. In the 'FM-$y$' state the spins point along the $y$-axis, perpendicular to the chain direction. In the 'FM-$xz$' state magnetic moments lie in the $x-z$ plane being canted out of the $x-y$ plane at an angle
\begin{align}
    \tan 2\theta=\frac{J_{z\pm}}{J_\text{iso}(1-\Delta)+J_{\pm\pm}},
    \label{eq_tilt}
\end{align}
which we also indicate in the phase diagram. These states are illustrated in Fig.~\ref{fig_chain_pd}. FM-$xz$ state is the one that is observed in experiments on \scgo \cite{ding2016}. Thus, when extracting parameters for \scgo\ we also check that the exchanges yield the correct ground state.

Using Holstein-Primakoff transformation
\begin{equation}
S^+_i=\sqrt{2S} a^{\phantom \dagger}_i, S^z_i=S-a^\dagger_i a^{\phantom \dagger}_i
\end{equation}
and standard spin-wave theory calculations we obtain energies at the $\mathbf{k}=0$ and $\mathbf{k}=(1/2,1/2,0)$, which can be expressed analytically:
\begin{align}
    \varepsilon_1 (0)&=\sqrt{\left(A-B\right)^2-\left(C+D+F\right)^2}\\
    \varepsilon_2 (0)&=\sqrt{\left(A+B\right)^2-\left(-C+D+F\right)^2}\nonumber\\
    \varepsilon_3 (0)&=\sqrt{\left(A+B\right)^2-\left(C+D-F\right)^2}\nonumber\\
    \varepsilon_4 (0)&=\sqrt{\left(A-B\right)^2-\left(C-D+F\right)^2}\nonumber
\end{align}
\begin{align}
    \varepsilon_1 (1/2,1/2,0)=\sqrt{\left(A-B\right)^2-D^2}\\
    \varepsilon_2 (1/2,1/2,0)=\sqrt{\left(A+B\right)^2-D^2}\nonumber
\end{align}
where
\begin{align}
    A&=-2\mathrm{J}^{zz}+4J_1+2J_2\\
    B&=-\left(\mathrm{J}^{xx}+\mathrm{J}^{yy}\right) \nonumber\\
    C&=4J_1 \nonumber\\
    D&=-\left(\mathrm{J}^{xx}-\mathrm{J}^{yy}\right)\nonumber\\
    F&=2J_2. \nonumber
\end{align}
The $\mathrm{J}^{\alpha\beta}$ are elements of the exchange matrix $\hat{\bm J}_{ij}$ in the local reference frame of the FM-$xz$ state, which contribute to linear spin-wave theory:
\begin{align}
\mathrm{J}^{xx}&=J_\text{iso}\left(\sin^2\theta+\Delta \cos^2\theta\right)+J_{\pm\pm} \sin^2 \theta-\frac{J_{z\pm}}{2}\sin 2\theta\nonumber\\
\mathrm{J}^{yy}&=J_\text{iso}-J_{\pm\pm}\nonumber\\
\mathrm{J}^{zz}&=J_\text{iso}\left(\cos^2\theta+\Delta \sin^2\theta\right)+J_{\pm\pm} \cos^2 \theta+\frac{J_{z\pm}}{2}\sin 2\theta.
\end{align}

The expressions above are used to extract exchanges using constraints:
\begin{align}
    &\varepsilon_1 (0)=1.0\text{ meV},\nonumber\\
    &\varepsilon_1 (1/2,1/2,0)=1.9\text{ meV},\nonumber\\
    &\varepsilon_2 (1/2,1/2,0)=2.7\text{ meV}.
\end{align}
Thus, for a fixed set of inter-chain exchanges $(J_1,J_2)$, the constraints above, together with the constraint on the tilt angle \eqref{eq_tilt} allows the extraction of the four exchanges of the model \eqref{HJpm}. Linear transformation \eqref{eq_ice_to_K} yields the exchanges of the Kitaev-Heisenberg model \eqref{eq_H_JKGGp}.

Curie-Weiss temperature can be calculated using the following expression \cite{valenti_CW}:
\begin{align}
\theta_\text{CW}=-\frac{S(S+1)}{3k_B}\left(4J_1+2J_2+\frac{2}{3}(2J_\text{iso}+\Delta J_\text{iso})\right).
\end{align}

\subsection*{Data availability}
The datasets generated during the current study are available under the following publicly accessible digital object identifiers (DOI) on the ONCat platform of Oak Ridge National Laboratory, https://doi.org/10.14461/ oncat.data. 6541161d1f3c8fcc291bc2df/ 2204019 and https://doi.org/10.14461/ oncat.data. 64d649ab9a4f9cddec92b19a/ 2204020 for the CNCS and SEQUOIA, respectively.

\subsection*{Code availability}
The code supporting this study's findings is available from the public GitHub repository: https://github.com/pavelmaxjinr/SrCoGe2O6

\section*{Acknowledgement}
The authors thank Jong Keum for assistance with X-ray measurements. We also would like to thank Sasha Chernyshev for useful discussions. Work at Oak Ridge National Laboratory (ORNL) was supported by the U.S. Department of Energy (DOE), Office of Science, Basic Energy Sciences, Materials Science and Engineering Division. This research used resources at the Spallation Neutron Source, a DOE Office of Science User Facility operated by the Oak Ridge National Laboratory. X-ray was conducted at the Center for Nanophase Materials Sciences (CNMS) (CNMS2019-R18) at ORNL, which is a DOE Office of Science User Facility. Magnetization measurements were conducted as part of a user project at the Center for Nanophase Materials Sciences (CNMS), which is a US Department of Energy, Office of Science User Facility at Oak Ridge National Laboratory.
Theoretical calculations of P.A.M., A.V.U., and S.V.S. were supported by the Russian Science Foundation via project RSF 23-12-00159, A.F.G thanks to program 122021000031-8 (Flux).


\bibliography{jpp_bib}

\end{document}